\documentclass[aps,prl,twocolumn,showpacs,superscriptaddress]{revtex4-1}

\usepackage{graphics}
\usepackage{graphicx}
\usepackage{dcolumn}
\usepackage{bm}
\usepackage{color}
\usepackage{amssymb,amsmath}

\begin{document}

\title{Ionization-induced asymmetric self-phase modulation and universal modulational instability in gas-filled hollow-core photonic crystal fibers}
\author{Mohammed F. Saleh}
\author{Wonkeun Chang}
\author{John C. Travers}
\affiliation{Max Planck Institute for the Science of Light, G\"{u}nther-Scharowsky str. 1, 91058 Erlangen, Germany}
\author{Philip St.J. Russell}
\affiliation{Max Planck Institute for the Science of Light, G\"{u}nther-Scharowsky str. 1, 91058 Erlangen, Germany}
\affiliation{Department of Physics, University of Erlangen-Nuremberg, Germany}
\author{Fabio Biancalana}
\affiliation{Max Planck Institute for the Science of Light, G\"{u}nther-Scharowsky str. 1, 91058 Erlangen, Germany}
\date{\today}

\begin{abstract}
We study theoretically the propagation of relatively long pulses with ionizing intensities in a hollow-core photonic crystal fiber filled with a Raman-inactive gas. Due to photoionization, previously unknown types of asymmetric self-phase modulation and `universal' modulational instabilities existing in both normal and anomalous dispersion regions appear. We also show that it is possible to spontaneously generate a plasma-induced continuum of blueshifting solitons, opening up new possibilities for pushing supercontinuum generation towards shorter and shorter wavelengths.
\end{abstract}
\maketitle

\paragraph{Introduction ---}
The invention of the photonic crystal fiber (PCF) has led to a true revolution in the field of nonlinear fiber optics \cite{Russell03}. Hollow-core PCFs (HC-PCFs) with the so-called kagome-lattice cladding have become in recent years a superior host for the investigation of light-matter interactions between intense ultrashort optical pulses and gaseous or liquid media \cite{Russell06,Travers11}. These fibers are characterized by a broadband transmission range with low group velocity dispersion (GVD), and a high confinement of light in the core.

In a series of recent groundbreaking experiments, few-$\mu$J femtosecond-scale pulses have been launched into argon-filled cm-long HC-PCFs, leading to the unique phenomenon of the photoionization-induced soliton self-frequency blueshift \cite{Hoelzer11b}. These experiments are well described by a unidirectional pulse propagation equation (UPPE) based on the full electric field, complemented by sophisticated tunneling and multiphoton ionization models \cite{Chang11,Kinsler10}. However, the UPPE model can be reduced to a set of two coupled equations for the envelope of the electric field and the ionization fraction \cite{Saleh11a,Saleh11b}, and remarkable qualitative agreement obtained.

In this Letter we show theoretically that when the gas is excited by relatively {\em long} pulses with ionizing intensities, new kinds of self-phase modulation (SPM) and modulational instability (MI) can emerge during propagation. Moreover, after the initial stage of instability is over, a `shower' of hundreds of solitons, each undergoing an ionization-induced self-frequency blueshift, pushes the supercontinuum spectrum towards shorter and shorter wavelengths. Such a {\em blueshifting} plasma-induced continuum has some similarities with the {\em redshifting} Raman-induced continuum driven by the Raman self-frequency shift in conventional solid-core fibers \cite{Islam89,GouveiaNeto89,Dianov89}.

\paragraph{Governing equations ---} The propagation of light in a HC-PCF filled with an ionizable Raman-inactive gas can be modeled by the following normalized coupled equations \cite{Saleh11a,Saleh11b}:
\begin{equation}
i\partial_{\xi}\psi+\hat{D}\left( i\partial_{\tau}\right)\psi+\left|\psi\right|^{2}\psi-\phi\,\psi+i\alpha\,\psi=0  ,\vspace{-5mm}\label{eq1}
\end{equation}
\begin{equation}
\partial_{\tau}\phi=\sigma(\phi_{\mathrm{T}}-\phi)\,\Delta|\psi|^{2}\,\Theta\left( \Delta|\psi|^{2}\right) -r\,\phi^{2} ,\label{eq2}
\end{equation}
where $\psi$ is the electric field envelope, $ \phi $ is a quantity proportional to the number of electrons created by the photoionization process, $\xi$ is the normalized longitudinal coordinate along the fiber, $\tau$ is time in a reference frame moving with the input pulse group velocity, $\hat{D}(i\partial_{\tau})$ is the full GVD operator, $ \alpha=\kappa\,\left(\phi_{\mathrm{T}}-\phi \right)\, \left[1-|\psi|^{2}_{\mathrm{th}}/|\psi|^{2} \right]\, \Theta\left( \Delta|\psi|^{2}\right) $ is the ionization-induced loss, $ \kappa $ is a normalization factor, (see also Refs. \cite{Agrawal07,Saleh11a,Saleh11b} for extensive details), $\Theta$ is the Heaviside step function, $\Delta|\psi|^{2}\equiv|\psi|^{2}-|\psi|^{2}_{\rm th}$,  $ |\psi|_{\mathrm{th}}^{2} $ is the normalized ionization threshold intensity, $ \sigma $ and $ \phi_{\mathrm{T}} $ are constants that can be determined through the tunneling ionization equation rate \cite{Saleh11a,Saleh11b}, and $ r $ is a coefficient regulating the recombination between free electrons and ions, which is a two-particle process and hence depends on $\phi^{2}$.

Equations (\ref{eq1}-\ref{eq2}) yield extraordinary qualitative agreement with experiments using ultrashort input pulses \cite{Hoelzer11b}, accurately predicting the dynamics of soliton formation and the plasma-induced soliton self-frequency blueshift \cite{Saleh11a,Saleh11b}. At the core of the model is the assumption that tunneling photoionization is dominant in the dynamics, and that the ionization rate can be described by a linear function corrected by a Heaviside function that takes into account the threshold intensity \cite{Saleh11a,Saleh11b,Travers11}. Although the long pulses considered here are expected to drive some multiphoton ionization, comparisons of Eqs. (\ref{eq1},\ref{eq2}) to UPPE simulations using the more general Yudin-Ivanov ionization model \cite{Yudin01} -- which we present at the end of this Letter -- show excellent agreement.

\paragraph{Plasma-induced asymmetric self-phase modulation ---} Ionization-induced SPM can be studied by simplifying Eqs. (\ref{eq1}-\ref{eq2}) for the case of small dispersion and long input pulse durations $t_{0}$. In this situation (natural for kagome-PCFs since the GVD is very small), the nonlinearity initially dominates over the GVD, since the `soliton number' $N$ (i.e. the input pulse energy) is large~\cite{Agrawal07}, and the second term in Eq. (\ref{eq1}) can be safely neglected in the very initial stage of propagation. In kagome-PCFs, input energies such that $N>100$ are realistically achievable for pulses with $ > $ 0.5 ps duration. Neglecting also the recombination process, Eq. (\ref{eq2}) can be written as $ \phi\left(\tau \right) =\phi_{\mathrm{T}}\,\left\lbrace 1-\exp\left[-\sigma\int_{-\infty}^{\tau}\Delta|\psi(\tau')|^{2}\,\Theta(\Delta|\psi(\tau')|^{2})\,d\tau'  \right]  \right\rbrace   $. For small values of $ \sigma $, Eqs. (\ref{eq1},\ref{eq2}) can be reduced to a single integro-differential equation,
\begin{equation}
i\partial_{\xi}\psi+\left|\psi\right|^{2}\psi-\eta\,\psi\!\!\!\int_{-\infty}^{\tau}\!\!\!\!\!\Delta|\psi(\tau')|^{2}\,\Theta(\Delta|\psi(\tau')|^{2})\,d\tau'+i\alpha\,\psi=0,\label{eq3}
\end{equation}
where $ \eta=\sigma\,\phi_{\mathrm{T}} $. By temporarily ignoring the losses (which do not change the qualitative picture that we are going to describe, but only saturate the SPM spectrum after a certain distance), Eq. (\ref{eq3}) can be solved by substituting $\psi\left(\xi,\tau \right) = V\left(\xi,\tau \right) \exp\left[\,\varphi_{\mathrm{NL}}\left(\xi,\tau \right)\right]  $ \cite{Agrawal07}, where $ V $ and $ \varphi_{\mathrm{NL}} $ are the amplitude and the nonlinear phase of the pulse. Equating the real and the imaginary parts, and performing the integrations results in $ V\left(\xi,\tau \right)=\psi\left(0,\tau \right) $, $ \varphi_{\mathrm{NL}}\left(\xi,\tau \right)=\xi\,\left[|V|^{2}- \int_{-\infty}^{\tau}\Delta|V(\tau')|^{2}\,\Theta(\Delta|V(\tau')|^{2})\,d\tau' \right]  $, and $ \Delta|V|^{2}= |V|^{2}-|\psi|^{2}_{\rm th}$. We will show below that such a nonlinear phase can induce a strong spectrally asymmetric SPM at a large rate compared to the background Kerr nonlinearity. For this purpose, we compute the mean frequency $ \left\langle \Omega\right\rangle  $, and the variance $   \left( \Delta\Omega \right) ^{2} = \left\langle\Omega^{2} \right\rangle -\left\langle\Omega \right\rangle^{2} $  \cite{Pinault85,Agrawal07}, where $ \left\langle\Omega^{k} \right\rangle = \int\Omega^{k} \left|\Psi\right|^{2} \, d\Omega/\int \left|\Psi\right|^{2} \, d\Omega$, $ \Psi=\mathcal{F}\left[ \psi\right]  $, and the symbol $ \mathcal{F} $ represents the Fourier transform. 

In the case of an input Gaussian pulse, $\psi\left(0,\tau \right) =\exp \left(-\tau^{2}/2\,\tau_{0}^{2} \right) $, the above parameters can be determined in closed forms. We find
\begin{equation}
\left\langle \Omega\right\rangle =\frac{1}{2}\,\eta\,\xi\left[ \sqrt{2}\,\mathrm{erf}\left(\sqrt{2}\,\mathcal{T} \right) - 2\,|\psi|^{2}_{\mathrm{th}}\,\mathrm{erf}\left(\mathcal{T}\right) \right],
\end{equation}
\begin{equation}
\begin{array}{ll}
 \left( \Delta\Omega \right) ^{2}=&\dfrac{1}{18\,\tau_{0}^{2}} \left\lbrace 9+ 4\sqrt{3}\,\xi^{2} -3\,\xi^{2}\eta^{2}\tau_{0}^{2}\left[ \,3\,\mathrm{erf}^{2}\left(\sqrt{2}\,\mathcal{T}\right)\right. \right.  \vspace{1mm} \\
 &    \left. -2\,\sqrt{3}\,\mathrm{erf} \left(\sqrt{3}\,\mathcal{T}\right)\right]+6|\psi|^{2}_{\mathrm{th}}\left[\, \mathrm{erf}\left(\mathcal{T}\right)-1\right] \vspace{1mm}\\
   & \left.  \times \left[\,|\psi|^{2}_{\mathrm{th}} \mathrm{erf}\left(\mathcal{T}\right)-\sqrt{2}\, \mathrm{erf}\left(\sqrt{2}\,\mathcal{T}\right)\right] \right\rbrace ,
\end{array}
\end{equation}
where erf is the error function, $ \mathcal{T}=\mathrm{T}/\tau_{0} $, and $\mathrm{-T} \leq \tau\leq \mathrm{T} $ is the regime within which the pulse intensity exceeds the threshold intensity.

\begin{figure}
\includegraphics[width=8.6cm]{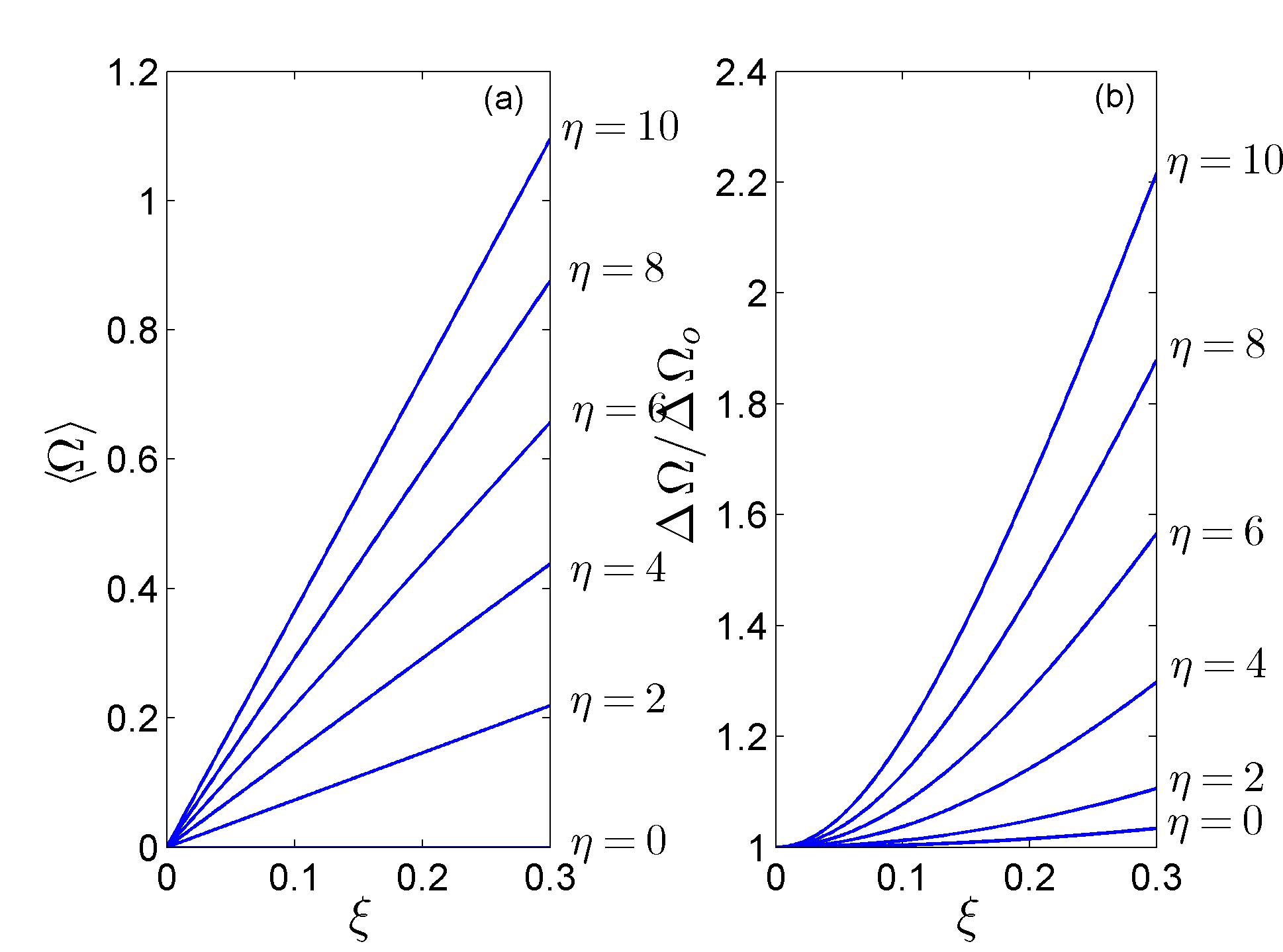}
\caption{(Color online). Spatial dependence of (a) the mean frequency $ \left\langle \Omega\right\rangle  $ and (b) the frequency standard deviation $ \Delta\Omega $ of a Gaussian pulse $  \exp \left(-\tau^{2}/2\,\tau_{0}^{2} \right) $ with $ \tau_{0}=2 $. The temporal position $ \mathrm{T}$ at which the pulse intensity can initiate photoionization is assumed to be equal to $\tau_{0} $. $ \Delta \Omega_{0} $ is the spectral width at $ \xi=0 $. \label{Fig1}}
\end{figure}

Panels (a,b) in Fig. \ref{Fig1} depict the spatial dependence of the mean frequency and the standard deviation, respectively, for different values of $ \eta $, i.e. for different free-electron densities generated in the fiber. In Fig. \ref{Fig1}(a), for $ \eta=0$, which corresponds to the absence of ionization, the mean frequency is always zero during propagation due to the well-known symmetric spectral broadening due to conventional SPM \cite{Agrawal07}. As $ \eta $ increases (i.e. when the plasma starts to build up inside the fiber), the mean frequency moves linearly towards the blue-side of the spectrum due to the ionization-induced phase-modulation. This induces a strong spectrally {\em asymmetric} SPM that is unique to the kind of gas-filled PCF waveguides studied in this Letter. By looking at Fig. \ref{Fig1}(b), we find that the ionization-induced SPM broadens the spectrum significantly faster than the well-known Kerr-induced SPM. The spectral broadening process is limited by the unavoidable ionization and fiber losses, similar to Kerr SPM.

\paragraph{Plasma-induced modulational instability ---} After the very initial SPM stage (described above) is over, the interplay between nonlinear and dispersive effects can lead to an instability that modulates the temporal profile of the pulse, creating new spectral sidebands referred to as modulational instability (MI) in Refs. \cite{Hasegawa80,Tai86a,Tai86b}). For instance, during the propagation of a continuous wave (CW) signal in an anomalously dispersive Kerr medium, perturbations in the amplitude of the carrier wave can trigger the generation of spectral sidebands, leading to the eventual break-up of the CW signal into a train of pulses \cite{Agrawal07,Hasegawa80,Tai86a}. MI due to the photoionization nonlinearity described by Eqs. (\ref{eq1}-\ref{eq2}) can be investigated by using the standard approach presented in Ref. \cite{Agrawal07,Hasegawa80}. The steady state solutions of Eqs. (\ref{eq1},\ref{eq2}), which are time independent, are found by inserting $ \psi=\psi_{0}\, e^{ik_{0}\xi} $ and $ \phi=\phi_{0} $. Perturbing these steady state solutions and substituting back into Eqs. (\ref{eq1},\ref{eq2}), one obtains the linearized equations
\begin{equation}
i\partial_{\xi}a+\frac{s}{2}\,\partial^{2}_{\tau}a+\psi_{0}^{2}\left(a+a^{*} \right) -u \psi_{0}=0\vspace{-5mm} ,\label{eq4}
\end{equation}
\begin{equation}
\partial_{\tau}u=-Fu+g\left(a+a^{*} \right),\label{eq5}
\end{equation}
where $ a $ and $ u $ are perturbations added to the amplitudes $ \psi_{0} $ and $ \phi_{0} $, $s$ is the sign of the second-order dispersion ($+1 \equiv$ anomalous dispersion, $-1 \equiv$ normal dispersion), higher-order dispersion coefficients are neglected, $ F= \sigma\left(\psi_{0}^{2}-|\psi|^{2}_{\mathrm{th}} \right)\,\Theta\,\left(\psi_{0}^{2}-|\psi|^{2}_{\mathrm{th}} \right)+2\,r\phi_{0}  $, $ g=\sigma(\phi_{\mathrm{T}}-\phi)\psi_{0} $. Assuming $ a=a_{1}\, e^{i\vartheta}+a_{2}\,e^{-i\vartheta^{*}}$, $ u=u_{0}\, e^{i\vartheta}+u_{0}^{*}\, e^{-i\vartheta^{*}}$ and $ \vartheta=\kappa\,\xi-\Omega\,\tau $, a set of two homogeneous equations for $ a_{1} $ and $ a_{2}^{*} $ can be found,
\begin{equation}
\left[ \begin{array}{cc}
m_{11}& m_{12} \\
 m_{21}&   m_{22}
\end{array}\right] \left[ \begin{array}{c}
a_{1} \\
a_{2}^{*}
\end{array}\right]   =0,
\end{equation}
where $ m_{11}= -\kappa-s\,\Omega^{2}/2+\psi_{0}^{2}+g\,\psi_{0}/\left( -F+i\,\Omega\right) $, $ m_{12}= \psi_{0}^{2}+g\,\psi_{0}/\left( -F+i\,\Omega\right)$, $ m_{21}=-m_{12} $, and $ m_{22}= -m_{11}-2\kappa $. This set has a nontrivial solution when
\begin{equation}
\kappa=\pm\frac{|\Omega|}{2}\,\left[\Omega^{2}-4s\left(\psi_{0}^{2}- \dfrac{g\,\psi_{0}\left( F+i\,\Omega\right) }{F^{2}+\Omega^{2}}\right) \right] ^{1/2},
\end{equation}
which are the complex eigenvalues of the problem. Note that in this analysis $ \phi $ is kept to be a real positive quantity since it represents the number of electrons generated via photoionization.

\begin{figure}
\includegraphics[width=8.6cm]{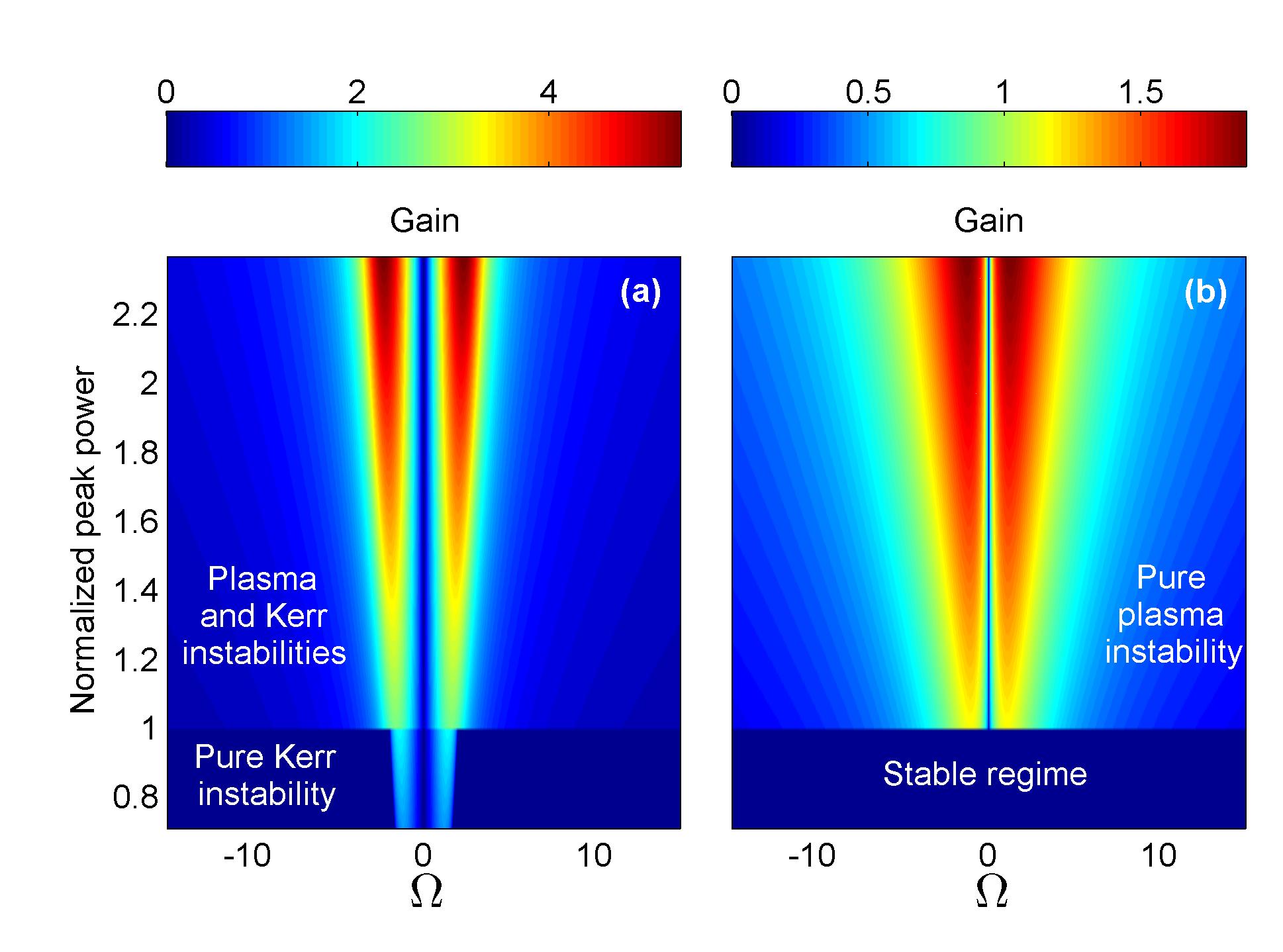}
\caption{(Color online). MI spectral gain profile versus normalized input peak power in (a) anomalous and (b) normal dispersion regimes with equal magnitude of GVD.
\label{Fig2}}
\end{figure}

Perturbations can only be amplified during propagation for frequencies which have non-real propagation constant $ \kappa $. The well-known Kerr-induced MI (e.g. in solid core optical fibers or in HC-PCFs when excited with intensities below the ionization threshold for the gas) occurs only in the anomalous dispersion regime \cite{Agrawal07}. However, the presence of the photoionization process induces a very unusual instability that can exist {\em in both normal and anomalous dispersion regimes, and for any frequency}. The reason for this `universal' instability is that the ionization contribution to the destabilization of the background wave $ g\,\psi_{0}/\left( -F+i\,\Omega\right) $ is always complex and frequency-dependent, in contrast to when only the Kerr effect is present. The spectral dependence of the gain, which is defined as $2\,\mathrm{Im} \left\lbrace \kappa\right\rbrace  $ for different peak powers, is shown in Fig. \ref{Fig2} for (a) anomalous and (b) normal dispersion regimes, where the physical powers are normalized to the threshold ionization power $ P_{\mathrm{th}} $, i.e., $ |\psi|^{2} _{\mathrm{th}} =1 $. When the normalized input power $ \psi_{0}^{2}\leq |\psi|^{2}_{\mathrm{th}} $, we have the traditional side-lobes, which exist uniquely in the anomalous regime, due to the Kerr-nonlinearity. However when $ \psi_{0}^{2} > |\psi|^{2}_{\mathrm{th}} $, photoionization-induced instability causes the side lobes to be unbounded and to have slowly-decaying tails. A similar situation occurs in the normal regime, where the gain is slightly lower due to the absence of the Kerr contribution. For this case there are no instabilities below the threshold power since no plasma is generated.

\paragraph{Numerical simulations ---}
To validate the analysis based on the simplified model given by Eqs. (\ref{eq1},\ref{eq2}), and also to provide further insight to the ionization driven MI dynamics, UPPE simulations were performed \cite{Chang11}. The ionization rate was calculated using the Yudin-Ivanov model \cite{Yudin01}, which accurately describes both quasi-static tunneling and multiphoton ionization along with the transition between these regimes~\cite{Yudin01,Gkortsas11}. With this model previous experimental results have been reproduced \cite{Hoelzer11b,Travers11}.

A Gaussian pulse at $1064$ nm with duration $1.18$ ps and peak power $200$ MW was launched into a kagome HC-PCF, core diameter $20$ $\mu$m, filled with argon gas to a pressure of 1 bar. This corresponds to anomalous dispersion with $\beta_{2} \simeq-2.8$ ps$ ^{2} $/km. In this fiber $P_{\mathrm{th}} \simeq105$~MW and the pump pulse has a soliton number $N\simeq102$. The peak of this pulse corresponds to a Keldysh parameter of $\gamma=0.785$, and for most of the pulse $\gamma<5$; this is the non-adiabatic tunneling regime~\cite{Yudin01}.

\begin{figure}
\includegraphics[width=8.6cm]{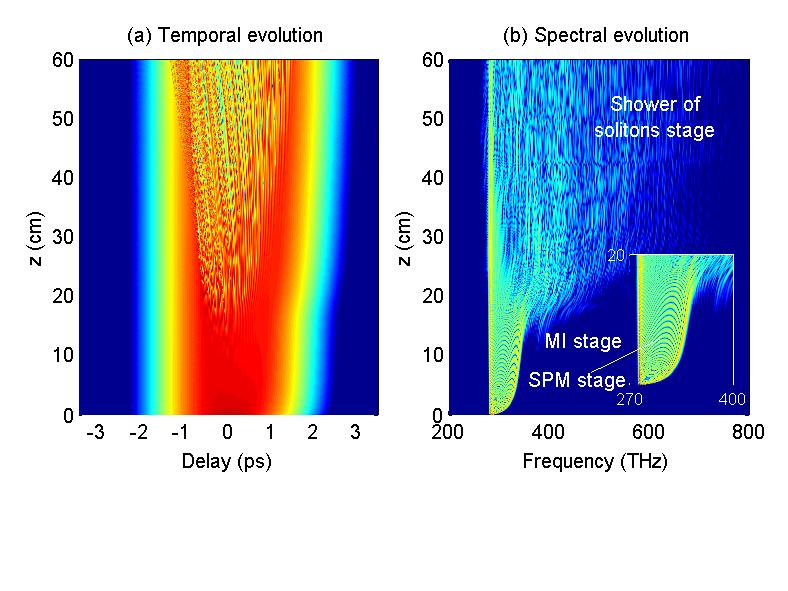}\vspace{-1.2cm}
\includegraphics[width=8.6cm]{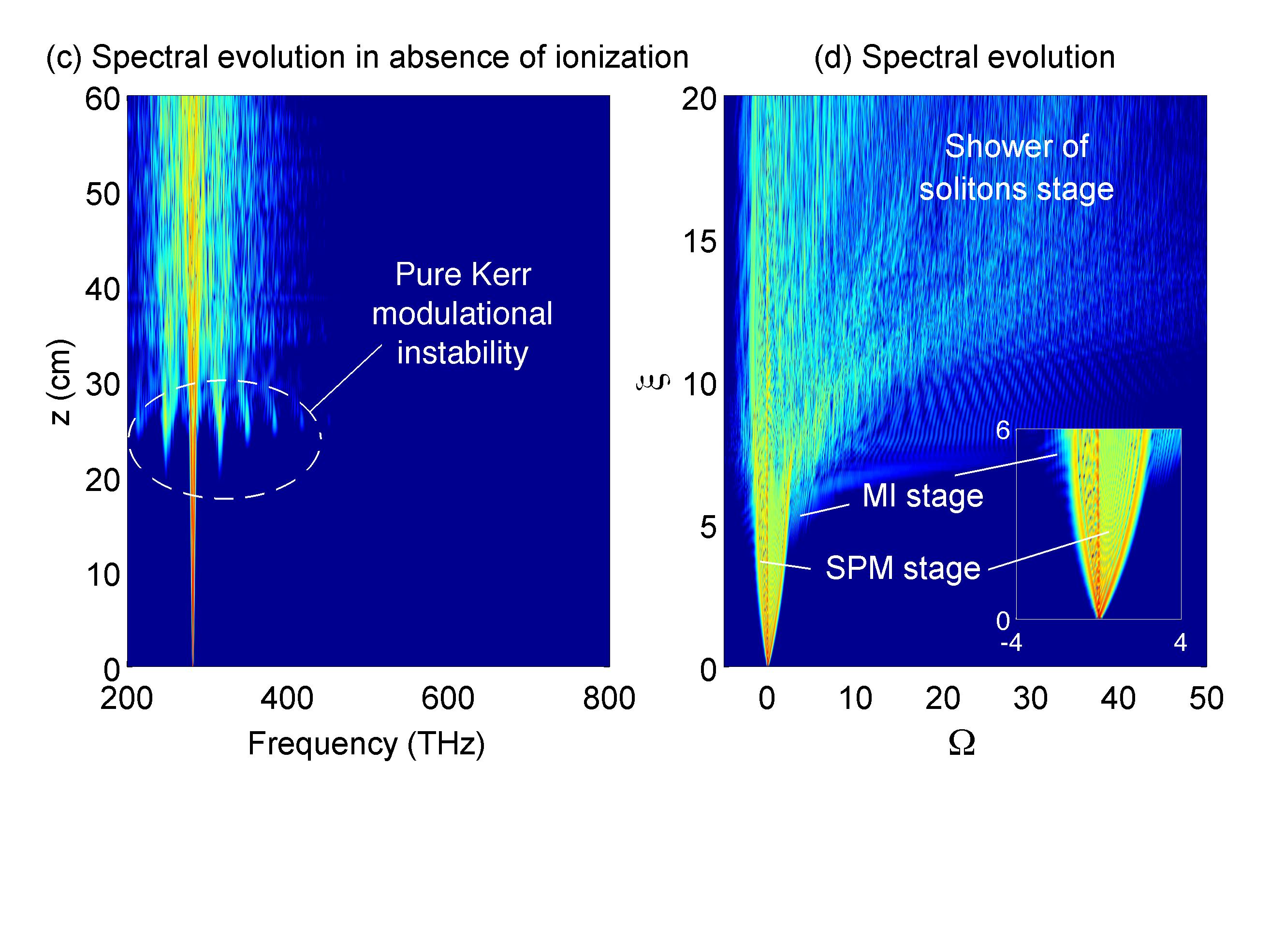}\vspace{-1.2cm}
\caption{(Color online). Spectral (a) and temporal (b) intensity evolution of a long Gaussian pulse propagating in an Ar-filled HC-PCF, calculated using a UPPE. (c) The corresponding spectral intensity evolution in absence of ionization. (d) The spectral intensity evolution obtained by solving coupled equations Eqs. (\ref{eq1},\ref{eq2}).
\label{Fig3}}
\end{figure}

Fig. \ref{Fig3}(a,b) shows the temporal and spectral evolution of the pulse through the fibre. The first stage of propagation shows asymmetric spectral broadening towards the blue due to ionization-induced SPM as described above -- see Fig. \ref{Fig3}(b) and inset. Immediately after the SPM stage, dispersion starts to play a role, and due to the combined Kerr and ionization MIs, unbounded side lobes are generated and amplified quickly -- see the oscillations in the time domain in Fig. \ref{Fig3}(a). In the third and final stage in the propagation, strongly blue-shifted solitons are emitted. For comparison, Fig. \ref{Fig3}(c) shows propagation without ionization. In this case weak side-lobes due to Kerr driven MI are visible, but no blue-asymmetry is evident. Fig. \ref{Fig3}(d) shows the spectral propagation obtained by solving the coupled equations Eqs. (\ref{eq1},\ref{eq2}). All of the qualitative features of the UPPE simulations are reproduced, validating this model and the subsequent analysis discussed in this Letter.

\begin{figure}
\includegraphics[width=8.6cm]{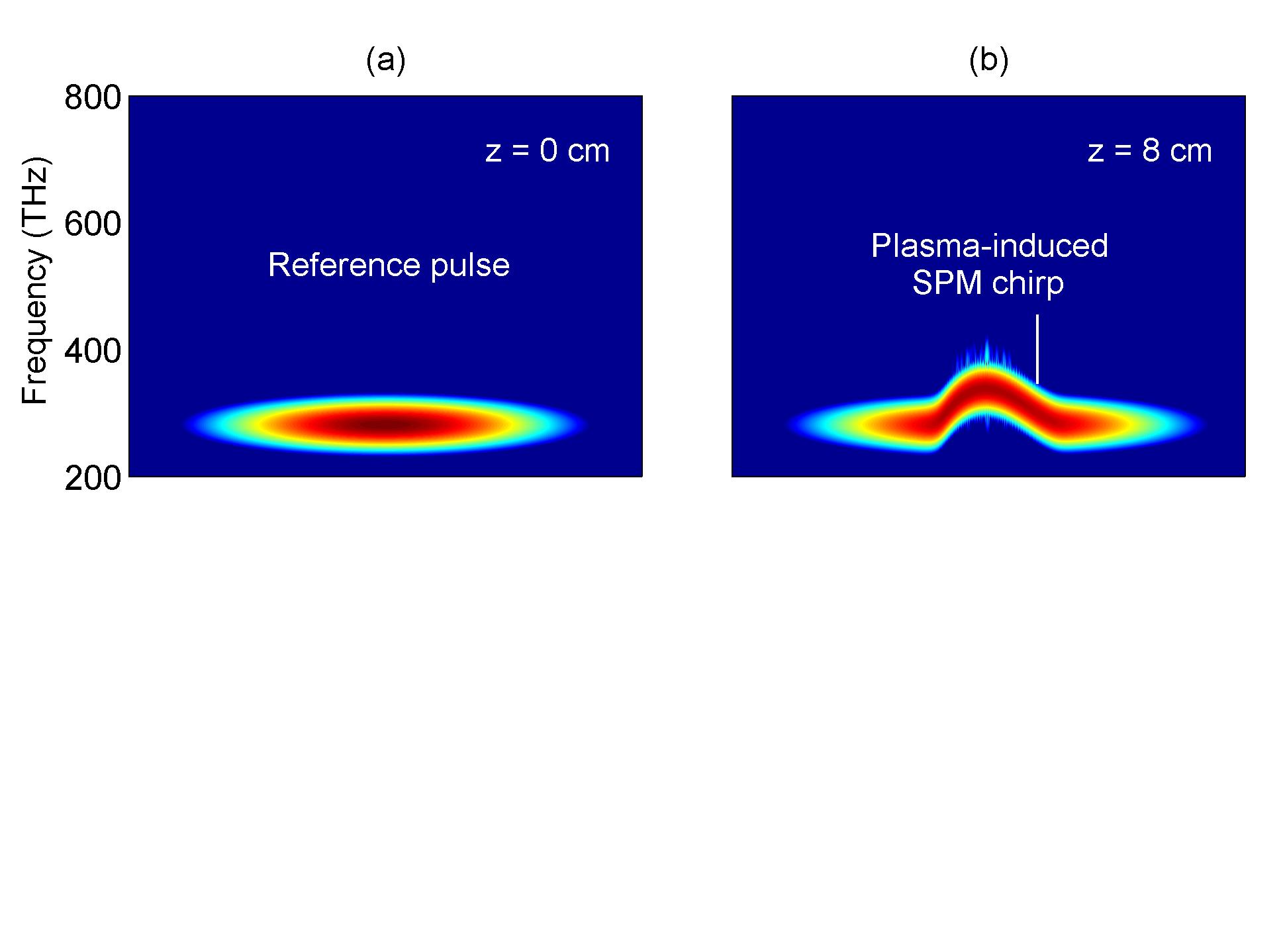}\vspace{-3cm}
\includegraphics[width=8.6cm]{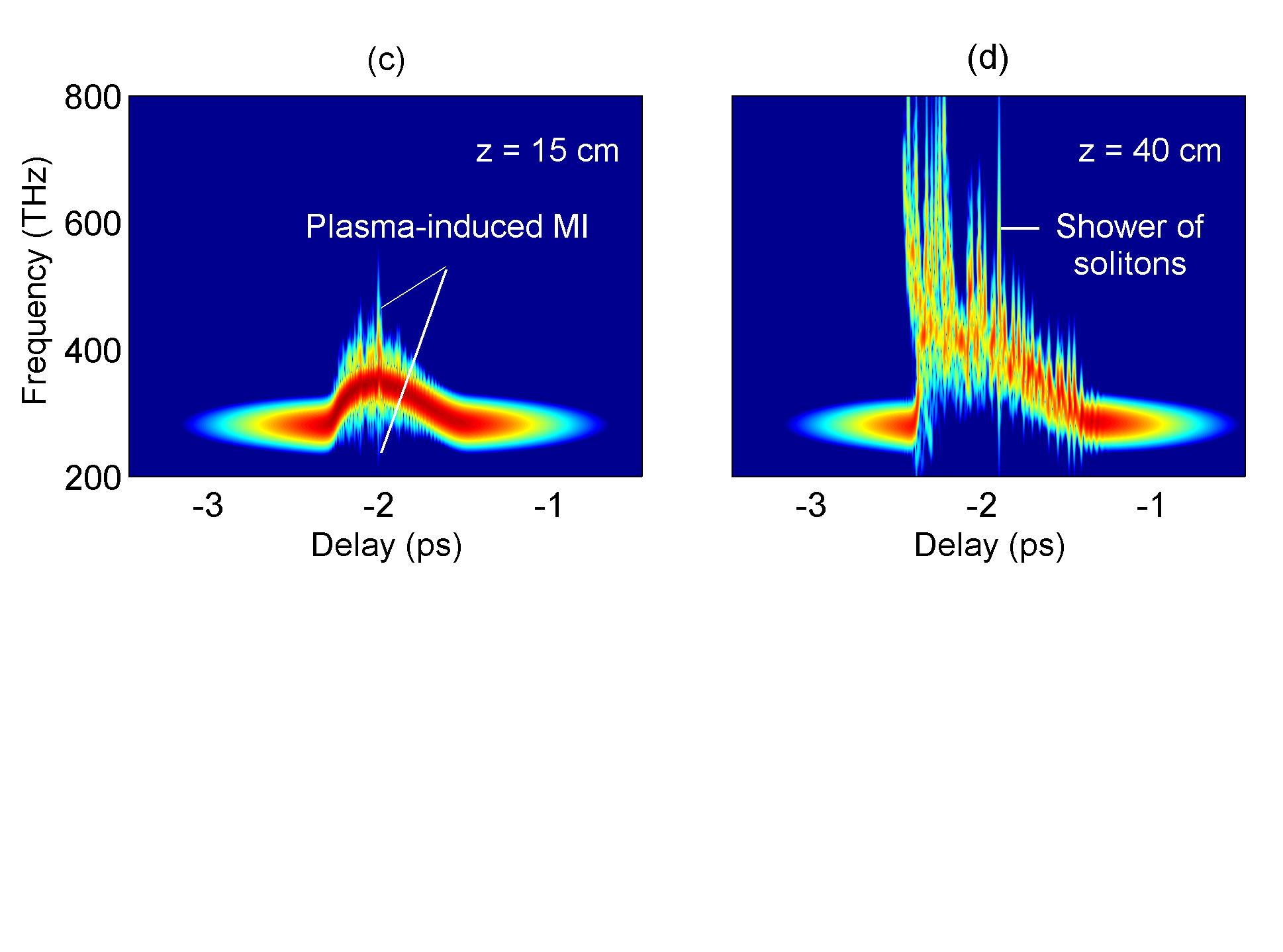}\vspace{-2.8cm}
\caption{(Color online). XFROG spectrograms for the propagation through an Ar-filled HC-PCF. The simulation parameters and color bars are the same as in Fig. \ref{Fig3}. (a) Reference pulse. (b) Frequency chirping and initiating MI. (c) Pulse break-up. (d) Soliton disintegration into multiple blueshifting solitons.
\label{Fig4}}
\end{figure}

Further insight into the dynamics can be obtained from Fig. \ref{Fig4}, where the evolution of the cross-frequency-resolved optical gating (XFROG) spectrograms of the pulse at different positions along the fiber is shown. The pulse is initially asymmetrically chirped to high frequencies in the centre of the pulse, [Fig. \ref{Fig4}(b)], due to the higher plasma density created at the peak intensities. This is the plasma-induced SPM chirp described above. At the same time two ionization-induced MI sidebands appear in the pulse spectrum [Fig. \ref{Fig4}(c)], as discussed above. MI facilitates the formation of many solitons. In less than half a meter of propagation the initial pulse disintegrates into a `shower' of solitary waves, see Figs. \ref{Fig4}(d), each undergoing a strong self-frequency blueshift induced by the intrapulse photoionization, as described in detail in Refs. \cite{Saleh11a,Saleh11b}.

\paragraph{Conclusions ---} In this Letter we have investigated analytically and numerically the propagation of intense and relatively long pulses in an Ar-filled kagome-cladding HC-PCF. Several surprising results have emerged, such as: strongly asymmetric SPM, which could be studied analytically with the envelope equations (\ref{eq1}-\ref{eq2}); a universal type of MI, existing in both anomalous and normal dispersion for all frequencies and exhibiting long tails in the gain spectra; and the final disintegration of the pulse into a multitude of blueshifting solitary waves, which form a  plasma-induced continuum. All these regimes are realistically accessible in experiments. These theoretical results further highlight the stimulating new possibilities opened up by accessing the ionization regime in hollow-core photonic crystal fibers.


\begin{thebibliography}{18}%
\makeatletter
\providecommand \@ifxundefined [1]{%
 \@ifx{#1\undefined}
}%
\providecommand \@ifnum [1]{%
 \ifnum #1\expandafter \@firstoftwo
 \else \expandafter \@secondoftwo
 \fi
}%
\providecommand \@ifx [1]{%
 \ifx #1\expandafter \@firstoftwo
 \else \expandafter \@secondoftwo
 \fi
}%
\providecommand \natexlab [1]{#1}%
\providecommand \enquote  [1]{``#1''}%
\providecommand \bibnamefont  [1]{#1}%
\providecommand \bibfnamefont [1]{#1}%
\providecommand \citenamefont [1]{#1}%
\providecommand \href@noop [0]{\@secondoftwo}%
\providecommand \href [0]{\begingroup \@sanitize@url \@href}%
\providecommand \@href[1]{\@@startlink{#1}\@@href}%
\providecommand \@@href[1]{\endgroup#1\@@endlink}%
\providecommand \@sanitize@url [0]{\catcode `\\12\catcode `\$12\catcode
  `\&12\catcode `\#12\catcode `\^12\catcode `\_12\catcode `\%12\relax}%
\providecommand \@@startlink[1]{}%
\providecommand \@@endlink[0]{}%
\providecommand \url  [0]{\begingroup\@sanitize@url \@url }%
\providecommand \@url [1]{\endgroup\@href {#1}{\urlprefix }}%
\providecommand \urlprefix  [0]{URL }%
\providecommand \Eprint [0]{\href }%
\providecommand \doibase [0]{http://dx.doi.org/}%
\providecommand \selectlanguage [0]{\@gobble}%
\providecommand \bibinfo  [0]{\@secondoftwo}%
\providecommand \bibfield  [0]{\@secondoftwo}%
\providecommand \translation [1]{[#1]}%
\providecommand \BibitemOpen [0]{}%
\providecommand \bibitemStop [0]{}%
\providecommand \bibitemNoStop [0]{.\EOS\space}%
\providecommand \EOS [0]{\spacefactor3000\relax}%
\providecommand \BibitemShut  [1]{\csname bibitem#1\endcsname}%
\let\auto@bib@innerbib\@empty
\bibitem [{\citenamefont {{P.~{St.J}.~Russell}}(2003)}]{Russell03}%
  \BibitemOpen
  \bibfield  {author} {\bibinfo {author} {\bibnamefont
  {{P.~{St.J}.~Russell}}},\ }\href@noop {} {\bibfield  {journal} {\bibinfo
  {journal} {Science}\ }\textbf {\bibinfo {volume} {299}},\ \bibinfo {pages}
  {358} (\bibinfo {year} {2003})}\BibitemShut {NoStop}%
\bibitem [{\citenamefont {{P.~{St.J}.~Russell}}(2006)}]{Russell06}%
  \BibitemOpen
  \bibfield  {author} {\bibinfo {author} {\bibnamefont
  {{P.~{St.J}.~Russell}}},\ }\href@noop {} {\bibfield  {journal} {\bibinfo
  {journal} {J. Lightwave Technol.}\ }\textbf {\bibinfo {volume} {24}},\
  \bibinfo {pages} {4729} (\bibinfo {year} {2006})}\BibitemShut {NoStop}%
\bibitem [{\citenamefont {Travers}\ \emph {et~al.}(2011)\citenamefont
  {Travers}, \citenamefont {Chang}, \citenamefont {Nold}, \citenamefont
  {Joly},\ and\ \citenamefont {{P.~{St.J}.~Russell}}}]{Travers11}%
  \BibitemOpen
  \bibfield  {author} {\bibinfo {author} {\bibfnamefont {J.~C.}\ \bibnamefont
  {Travers}}, \bibinfo {author} {\bibfnamefont {W.}~\bibnamefont {Chang}},
  \bibinfo {author} {\bibfnamefont {J.}~\bibnamefont {Nold}}, \bibinfo {author}
  {\bibfnamefont {N.~Y.}\ \bibnamefont {Joly}}, \ and\ \bibinfo {author}
  {\bibnamefont {{P.~{St.J}.~Russell}}},\ }\href@noop {} {\bibfield  {journal}
  {\bibinfo  {journal} {J. Opt. Soc. Am. B}\ }\textbf {\bibinfo {volume}
  {28}},\ \bibinfo {pages} {A11} (\bibinfo {year} {2011})}\BibitemShut
  {NoStop}%
\bibitem [{\citenamefont {H\"{o}lzer}\ \emph {et~al.}(2011)\citenamefont
  {H\"{o}lzer}, \citenamefont {Chang}, \citenamefont {Travers}, \citenamefont
  {Nazarkin}, \citenamefont {Nold}, \citenamefont {Joly}, \citenamefont
  {Saleh}, \citenamefont {Biancalana},\ and\ \citenamefont
  {{P.~{St.J}.~Russell}}}]{Hoelzer11b}%
  \BibitemOpen
  \bibfield  {author} {\bibinfo {author} {\bibfnamefont {P.}~\bibnamefont
  {H\"{o}lzer}}, \bibinfo {author} {\bibfnamefont {W.}~\bibnamefont {Chang}},
  \bibinfo {author} {\bibfnamefont {J.~C.}\ \bibnamefont {Travers}}, \bibinfo
  {author} {\bibfnamefont {A.}~\bibnamefont {Nazarkin}}, \bibinfo {author}
  {\bibfnamefont {J.}~\bibnamefont {Nold}}, \bibinfo {author} {\bibfnamefont
  {N.~Y.}\ \bibnamefont {Joly}}, \bibinfo {author} {\bibfnamefont {M.~F.}\
  \bibnamefont {Saleh}}, \bibinfo {author} {\bibfnamefont {F.}~\bibnamefont
  {Biancalana}}, \ and\ \bibinfo {author} {\bibnamefont
  {{P.~{St.J}.~Russell}}},\ }\href@noop {} {\bibfield  {journal} {\bibinfo
  {journal} {Phys. Rev. Lett.}\ }\textbf {\bibinfo {volume} {107}},\ \bibinfo
  {pages} {203901} (\bibinfo {year} {2011})}\BibitemShut {NoStop}%
\bibitem [{\citenamefont {Chang}\ \emph {et~al.}(2011)\citenamefont {Chang},
  \citenamefont {Nazarkin}, \citenamefont {Travers}, \citenamefont {Nold},
  \citenamefont {H\"{o}lzer}, \citenamefont {Joly},\ and\ \citenamefont
  {{P.~{St.J}.~Russell}}}]{Chang11}%
  \BibitemOpen
  \bibfield  {author} {\bibinfo {author} {\bibfnamefont {W.}~\bibnamefont
  {Chang}}, \bibinfo {author} {\bibfnamefont {A.}~\bibnamefont {Nazarkin}},
  \bibinfo {author} {\bibfnamefont {J.~C.}\ \bibnamefont {Travers}}, \bibinfo
  {author} {\bibfnamefont {J.}~\bibnamefont {Nold}}, \bibinfo {author}
  {\bibfnamefont {P.}~\bibnamefont {H\"{o}lzer}}, \bibinfo {author}
  {\bibfnamefont {N.~Y.}\ \bibnamefont {Joly}}, \ and\ \bibinfo {author}
  {\bibnamefont {{P.~{St.J}.~Russell}}},\ }\href@noop {} {\bibfield  {journal}
  {\bibinfo  {journal} {Opt. Express}\ }\textbf {\bibinfo {volume} {19}},\
  \bibinfo {pages} {21018} (\bibinfo {year} {2011})}\BibitemShut {NoStop}%
\bibitem [{\citenamefont {Kinsler}(2010)}]{Kinsler10}%
  \BibitemOpen
  \bibfield  {author} {\bibinfo {author} {\bibfnamefont {P.}~\bibnamefont
  {Kinsler}},\ }\href@noop {} {\bibfield  {journal} {\bibinfo  {journal} {Phys.
  Rev. A}\ }\textbf {\bibinfo {volume} {81}},\ \bibinfo {pages} {013819}
  (\bibinfo {year} {2010})}\BibitemShut {NoStop}%
\bibitem [{\citenamefont {Saleh}\ \emph {et~al.}(2011)\citenamefont {Saleh},
  \citenamefont {Chang}, \citenamefont {H\"olzer}, \citenamefont {Nazarkin},
  \citenamefont {Travers}, \citenamefont {Joly}, \citenamefont
  {{P.~{St.J}.~Russell}},\ and\ \citenamefont {Biancalana}}]{Saleh11a}%
  \BibitemOpen
  \bibfield  {author} {\bibinfo {author} {\bibfnamefont {M.~F.}\ \bibnamefont
  {Saleh}}, \bibinfo {author} {\bibfnamefont {W.}~\bibnamefont {Chang}},
  \bibinfo {author} {\bibfnamefont {P.}~\bibnamefont {H\"olzer}}, \bibinfo
  {author} {\bibfnamefont {A.}~\bibnamefont {Nazarkin}}, \bibinfo {author}
  {\bibfnamefont {J.~C.}\ \bibnamefont {Travers}}, \bibinfo {author}
  {\bibfnamefont {N.~Y.}\ \bibnamefont {Joly}}, \bibinfo {author} {\bibnamefont
  {{P.~{St.J}.~Russell}}}, \ and\ \bibinfo {author} {\bibfnamefont
  {F.}~\bibnamefont {Biancalana}},\ }\href@noop {} {\bibfield  {journal}
  {\bibinfo  {journal} {Phys. Rev. Lett.}\ }\textbf {\bibinfo {volume} {107}},\
  \bibinfo {pages} {203902} (\bibinfo {year} {2011})}\BibitemShut {NoStop}%
\bibitem [{\citenamefont {Saleh}\ and\ \citenamefont
  {Biancalana}(2011)}]{Saleh11b}%
  \BibitemOpen
  \bibfield  {author} {\bibinfo {author} {\bibfnamefont {M.~F.}\ \bibnamefont
  {Saleh}}\ and\ \bibinfo {author} {\bibfnamefont {F.}~\bibnamefont
  {Biancalana}},\ }\href@noop {} {\bibfield  {journal} {\bibinfo  {journal}
  {Phys. Rev. A}\ }\textbf {\bibinfo {volume} {84}},\ \bibinfo {pages} {063838}
  (\bibinfo {year} {2011})}\BibitemShut {NoStop}%
\bibitem [{\citenamefont {Islam}\ \emph {et~al.}(1989)\citenamefont {Islam},
  \citenamefont {Sucha}, \citenamefont {Bar-Joseph}, \citenamefont {Wegener},
  \citenamefont {Gordon},\ and\ \citenamefont {Chemla}}]{Islam89}%
  \BibitemOpen
  \bibfield  {author} {\bibinfo {author} {\bibfnamefont {M.~N.}\ \bibnamefont
  {Islam}}, \bibinfo {author} {\bibfnamefont {G.}~\bibnamefont {Sucha}},
  \bibinfo {author} {\bibfnamefont {I.}~\bibnamefont {Bar-Joseph}}, \bibinfo
  {author} {\bibfnamefont {M.}~\bibnamefont {Wegener}}, \bibinfo {author}
  {\bibfnamefont {J.~P.}\ \bibnamefont {Gordon}}, \ and\ \bibinfo {author}
  {\bibfnamefont {D.~S.}\ \bibnamefont {Chemla}},\ }\href@noop {} {\bibfield
  {journal} {\bibinfo  {journal} {Opt. Lett}\ }\textbf {\bibinfo {volume}
  {14}},\ \bibinfo {pages} {370} (\bibinfo {year} {1989})}\BibitemShut
  {NoStop}%
\bibitem [{\citenamefont {Gouveia-Neto}\ \emph {et~al.}()\citenamefont
  {Gouveia-Neto}, \citenamefont {Faldon},\ and\ \citenamefont
  {Taylor}}]{GouveiaNeto89}%
  \BibitemOpen
  \bibfield  {author} {\bibinfo {author} {\bibfnamefont {A.~S.}\ \bibnamefont
  {Gouveia-Neto}}, \bibinfo {author} {\bibfnamefont {M.~E.}\ \bibnamefont
  {Faldon}}, \ and\ \bibinfo {author} {\bibfnamefont {J.~R.}\ \bibnamefont
  {Taylor}},\ }\href@noop {} {\ }\BibitemShut {NoStop}%
\bibitem [{\citenamefont {Dianov}\ \emph {et~al.}(1989)\citenamefont {Dianov},
  \citenamefont {Grudinin}, \citenamefont {Khaidarov}, \citenamefont
  {Korobkin}, \citenamefont {Prokhorov},\ and\ \citenamefont
  {Serkin}}]{Dianov89}%
  \BibitemOpen
  \bibfield  {author} {\bibinfo {author} {\bibfnamefont {E.~M.}\ \bibnamefont
  {Dianov}}, \bibinfo {author} {\bibfnamefont {A.~B.}\ \bibnamefont
  {Grudinin}}, \bibinfo {author} {\bibfnamefont {D.~V.}\ \bibnamefont
  {Khaidarov}}, \bibinfo {author} {\bibfnamefont {D.~V.}\ \bibnamefont
  {Korobkin}}, \bibinfo {author} {\bibfnamefont {A.~M.}\ \bibnamefont
  {Prokhorov}}, \ and\ \bibinfo {author} {\bibfnamefont {V.~N.}\ \bibnamefont
  {Serkin}},\ }\href@noop {} {\bibfield  {journal} {\bibinfo  {journal} {Fiber
  and Integrated Optics}\ }\textbf {\bibinfo {volume} {8}},\ \bibinfo {pages}
  {61} (\bibinfo {year} {1989})}\BibitemShut {NoStop}%
\bibitem [{\citenamefont {Agrawal}(2007)}]{Agrawal07}%
  \BibitemOpen
  \bibfield  {author} {\bibinfo {author} {\bibfnamefont {G.~P.}\ \bibnamefont
  {Agrawal}},\ }\href@noop {} {\emph {\bibinfo {title} {Nonlinear Fiber
  Optics}}},\ \bibinfo {edition} {4th}\ ed.,\ San Diego, California\ (\bibinfo
  {publisher} {Academic Press},\ \bibinfo {year} {2007})\BibitemShut {NoStop}%
\bibitem [{\citenamefont {Yudin}\ and\ \citenamefont {Ivanov}(2001)}]{Yudin01}%
  \BibitemOpen
  \bibfield  {author} {\bibinfo {author} {\bibfnamefont {G.~L.}\ \bibnamefont
  {Yudin}}\ and\ \bibinfo {author} {\bibfnamefont {M.~Y.}\ \bibnamefont
  {Ivanov}},\ }\href@noop {} {\bibfield  {journal} {\bibinfo  {journal} {Phys.
  Rev. A}\ }\textbf {\bibinfo {volume} {64}},\ \bibinfo {pages} {013409}
  (\bibinfo {year} {2001})}\BibitemShut {NoStop}%
\bibitem [{\citenamefont {Pinault}\ and\ \citenamefont
  {Potasek}(1985)}]{Pinault85}%
  \BibitemOpen
  \bibfield  {author} {\bibinfo {author} {\bibfnamefont {S.~C.}\ \bibnamefont
  {Pinault}}\ and\ \bibinfo {author} {\bibfnamefont {M.~J.}\ \bibnamefont
  {Potasek}},\ }\href@noop {} {\bibfield  {journal} {\bibinfo  {journal} {J.
  Opt. Soc. Am. B}\ }\textbf {\bibinfo {volume} {2}},\ \bibinfo {pages} {1318}
  (\bibinfo {year} {1985})}\BibitemShut {NoStop}%
\bibitem [{\citenamefont {Hasegawa}\ and\ \citenamefont
  {Brinkman}()}]{Hasegawa80}%
  \BibitemOpen
  \bibfield  {author} {\bibinfo {author} {\bibfnamefont {A.}~\bibnamefont
  {Hasegawa}}\ and\ \bibinfo {author} {\bibfnamefont {W.}~\bibnamefont
  {Brinkman}},\ }\href@noop {} {\bibinfo  {journal} {IEEE. J. Quantum
  Electron.}\ }\BibitemShut {NoStop}%
\bibitem [{\citenamefont {Tai}\ \emph {et~al.}(1986{\natexlab{a}})\citenamefont
  {Tai}, \citenamefont {Hasegawa},\ and\ \citenamefont {Tomita}}]{Tai86a}%
  \BibitemOpen
\bibfield  {journal} {  }\bibfield  {author} {\bibinfo {author} {\bibfnamefont
  {K.}~\bibnamefont {Tai}}, \bibinfo {author} {\bibfnamefont {A.}~\bibnamefont
  {Hasegawa}}, \ and\ \bibinfo {author} {\bibfnamefont {A.}~\bibnamefont
  {Tomita}},\ }\href@noop {} {\bibfield  {journal} {\bibinfo  {journal} {Phys.
  Rev. Lett.}\ }\textbf {\bibinfo {volume} {56}},\ \bibinfo {pages} {135}
  (\bibinfo {year} {1986}{\natexlab{a}})}\BibitemShut {NoStop}%
\bibitem [{\citenamefont {Tai}\ \emph {et~al.}(1986{\natexlab{b}})\citenamefont
  {Tai}, \citenamefont {Tomita}, \citenamefont {Jewell},\ and\ \citenamefont
  {Hasegawa}}]{Tai86b}%
  \BibitemOpen
  \bibfield  {author} {\bibinfo {author} {\bibfnamefont {K.}~\bibnamefont
  {Tai}}, \bibinfo {author} {\bibfnamefont {A.}~\bibnamefont {Tomita}},
  \bibinfo {author} {\bibfnamefont {J.~L.}\ \bibnamefont {Jewell}}, \ and\
  \bibinfo {author} {\bibfnamefont {A.}~\bibnamefont {Hasegawa}},\ }\href@noop
  {} {\bibfield  {journal} {\bibinfo  {journal} {Appl. Phys. Lett.}\ }\textbf
  {\bibinfo {volume} {49}},\ \bibinfo {pages} {236} (\bibinfo {year}
  {1986}{\natexlab{b}})}\BibitemShut {NoStop}%
\bibitem [{\citenamefont {Gkortsas}\ \emph {et~al.}(2011)\citenamefont
  {Gkortsas}, \citenamefont {Bhardwaj}, \citenamefont {Lai}, \citenamefont
  {Hong}, \citenamefont {{Falc{\~a}o-Filho}},\ and\ \citenamefont
  {K\"artner}}]{Gkortsas11}%
  \BibitemOpen
  \bibfield  {author} {\bibinfo {author} {\bibfnamefont {V.~M.}\ \bibnamefont
  {Gkortsas}}, \bibinfo {author} {\bibfnamefont {S.}~\bibnamefont {Bhardwaj}},
  \bibinfo {author} {\bibfnamefont {C.~J.}\ \bibnamefont {Lai}}, \bibinfo
  {author} {\bibfnamefont {K.~H.}\ \bibnamefont {Hong}}, \bibinfo {author}
  {\bibfnamefont {E.~L.}\ \bibnamefont {{Falc{\~a}o-Filho}}}, \ and\ \bibinfo
  {author} {\bibfnamefont {F.~X.}\ \bibnamefont {K\"artner}},\ }\href@noop {}
  {\bibfield  {journal} {\bibinfo  {journal} {Phys. Rev. A}\ }\textbf {\bibinfo
  {volume} {84}},\ \bibinfo {pages} {013427} (\bibinfo {year}
  {2011})}\BibitemShut {NoStop}%
\end{thebibliography}

%

\end{document}